# Skyrmion confinement and dynamics in tracks patterned with magnetic anisotropy: theory and simulations

E. Tamura[*,1,2,#,†], C. Liu[*,3], S. Miki[*,1,2], J. Cho[1,2,4], M. Goto[1,2], H. Nomura[1,2,3], R. Nakatani[3], & Y. Suzuki[1,2#]

[1]Graduate School of Engineering Science, Osaka University, Toyonaka, Osaka 560-8531, Japan

[2]Center of Spintronics Research Network (CSRN), Graduate School of Engineering Science, Osaka University, Osaka, 560-8531, Japan

[3]Graduate School of Engineering, Osaka University, Suita, Osaka 565-0871, Japan

[4]Division of Nanotechnology, Daegu Gyeongbuk Institute of Science and Technology (DGIST), Daegu, 42988, Republic of Korea

* These three authors contributed equally to this work.
# Corresponding authors. suzuki-y@mp.es.osaka-u.ac.jp, tamura@spin.mp.es.osaka-u.ac.jp
† Present address: Department of Electronic Science and Engineering, Kyoto University, Kyoto, Kyoto, 615-8510, Japan

**Abstract**

Skyrmion is a topologically protected spin texture excited in magnetic thin films. The radii of skyrmions are typically 10〜100 nm. Because of the size, the skyrmion is expected to be a candidate for memory and novel-device usages. To realize the futuristic devices that will be using the skyrmion circuit, the tracks which guide the motion of skyrmions are needed. The tracks patterned with differences in the magnetic-anisotropy energy are well-paved without a potential pocket, whereas the tracks carved out of magnetic films have the potential pockets at corners due to the demagnetizing field. Therefore, the tracks patterned with the magnetic anisotropy plays a key role in making the skyrmion circuits. The experiment along this idea has been conducted for the hub and bent tracks. However, we have little known the motion of skyrmions in these tracks. This work aims to identify the forces acting between skyrmions and walls of the tracks.

The static force on a skyrmion can be expressed as minus the gradient of the potential energy caused by the magnetic-anisotropy undulation. The potential can be estimated numerically, modeling the shape of skyrmions with their radii and domain wall widths. We find that the forces depend not only on the distance from the wall but also on the shape of skyrmions. We have also performed micromagnetic simulations where the Magnus force and the acceleration by the magnetic-anisotropy gradient are taken into account as well as the force by the walls. The simulation results show good agreement with those calculated from the modeled skyrmions.

# Introduction

Skyrmion was initially proposed as a model of nuclear particles by Tony Skyrme in 1962 [1]. He successfully constructed a stable localized field configuration in continuous fields, which can be interpreted as a particle. Regarding a two dimensional magnetic thin film as Skyrme's continuous field, we have a localized spin configuration, that is, a skyrmion of the solid-state version and called the magnetic skyrmion [2].

Apart from a theoretical interest, the magnetic skyrmion offers great promise as an information carrier for memories [3] and devices with new logic [4] owing to its prominent features, including small size, topological stability, and the low critical current to move it. To realize these applications, the track which guides the motion of skyrmions will be an essential ingredient for constructing futuristic devices. A few pioneering experiments have been reported on the tracks patterned with the magnetic-anisotropy energy [5] and on controlling it by external electric fields [6]. These works have aimed to establish the necessary technology on target to the Brownian computation and shown that the thermal energy can drive skyrmions in their tracks. They have demonstrated how smooth their tracks are without the obstructive effects of demagnetization fields.

In the present work, we study the mechanism of the skyrmion confinement and dynamics in the tracks theoretically and identify the forces acting between skyrmions and walls of the tracks. We derive the equation of motion in terms of Thiele's manipulation [7,8] under the assumption of skyrmion shape [9]. Micromagnetic simulations [10,11] are also conducted to confirm the theoretical results. We performed the simulations under various conditions, including the effects of anisotropy gradient [12-14] and skyrmion mass [16-19].

## Results

The dynamics of the skyrmion are expressed by the Thiele equation [7]

$$\mathbf{0} = -\alpha D \dot{\mathbf{X}} + \mathbf{G} \times \dot{\mathbf{X}} + \mathbf{F}, \qquad (1)$$

where $\alpha$ is the damping constant, $D$ dissipation dyadic, $\mathbf{G}$ gyrocoupling vector, $\mathbf{F}$ external force for the potential, $\mathbf{X}$ coordinate of the skyrmion. $D$ depends on the shape of the skyrmion [8,9], i.e., $R/w$. $R$ is the radius, and $w$ is the domain wall width. $\mathbf{G}$ is proportional to the skyrmion number, $q$. In our theory, skyrmion with $q=1$ is treated. The extension of the theory to the cases with different $q$ is straight forward.

The potential energy of a skyrmion in the media with inhomogeneous magnetic anisotropy is expressed as follows,

$$U(\mathbf{X}) = \iint d^2\mathbf{r} K(\mathbf{r}) \left(1 - n_z^2(\mathbf{r};\mathbf{X})\right), \qquad (2)$$

where $K$ is the position-dependent magnetic anisotropy energy coefficient, $n_z$ the direction cosine of the magnetic vector field with a skyrmion at position $\mathbf{X}$.

In this paper, the interaction force between a skyrmion and a region with higher magnetic anisotropy (anisotropy-wall) will be discussed. In Fig. 1, a schematic of the system is drawn. In the anisotropy-wall, the magnetic anisotropy is $K_-$, and that in the track is $K_+$. Taking $K_- > K_+$ the skyrmion has lower energy in the track. If the skyrmion with the radius $R$ approach to the anisotropy-wall, the energy of the system (Eq. (2)) increases. The increase in the energy can be calculated as a function of skyrmion position $\mathbf{X}$, skyrmion radius $R$, and its wall width $w$ by assuming the rigid shape of the skyrmion. Then, $\mathbf{X}$-derivative of the energy provides the force as follows:

$$(\mathbf{F}_{\text{wall}})_x = (K_- - K_+) \int_{-\infty}^{\infty} dy \frac{4\sinh^2\left(\frac{R}{w}\right)\sinh^2\left(\frac{\sqrt{X^2+y^2}}{w}\right)}{\left(\sinh^2\left(\frac{R}{w}\right) + \sinh^2\left(\frac{\sqrt{X^2+y^2}}{w}\right)\right)^2}. \qquad (3)$$

In order to verify the rigid-shape assumption, a micro-magnetic simulation was done as a comparison. In Fig. 2, the theoretical force for skyrmions with different $R/w$ is displayed by solid curves. The simulation results are shown by solid circles and diamonds. The results indicated by the circles are obtained by adding a small slope in the anisotropy for the track region. The slope is parallel to the wall-track boundary ($y$-direction). By the slope, the skyrmion shows constant speed motion parallel to the boundary. The balance between a repulsive force from the anisotropy-wall and Magnus force because of finite velocity provides a measure of the interaction force between the skyrmion and the anisotropy-wall. The results indicated by the diamonds are obtained by adding a small slope in the anisotropy for the track region. The slope is perpendicular to the wall-track boundary. The balance between a repulsive force from the anisotropy-wall and force from an anisotropy slope provides a measure of the interaction force between the skyrmion and the anisotropy-wall. The simulation agrees well with those obtained from Eq. (3).

**Discussion**

In our model, we assume that isolated skyrmions do not alter their shapes described by their radii and domain wall widths as well as their centrosymmetric structures, during motion and under forces. Despite such a crude model, the numerically estimated forces are almost identical to those obtained from the simulations. It means that the difference in the potential energy causes the force rather than the deformation of skyrmion like a bouncing rubber ball. The force has a characteristic **X**-dependence where a peak locates at the radius of skyrmions since the force stems mostly from the transition part of the magnetization (the domain wall) overlapping the potential jump at **X=0**. Furthermore, it is symmetric for changing the sign of **X** that we can see in Eq. (3). The force also acts on the skyrmions that have crossed over the wall to push back them to tracks. The property might be useful to stabilize the skyrmion circuits. Another helpful feature is that the magnetic anisotropy wall does not give rise to additional friction. We found that the repulsive forces acting on skyrmions are identical whenever they are moving or at a stop.

## Method

The normalized magnetization **n** can be written in the spherical coordinate as follows,

$$\mathbf{n}(\Theta, \Phi) = \begin{pmatrix} \sin\Theta\cos\Phi \\ \sin\Theta\sin\Phi \\ \cos\Theta \end{pmatrix}. \quad (4)$$

For the skyrmion case, the azimuthal angle $\Theta$ depends on the skyrmion radius $R$ and domain wall width $w$ and the polar angle $\Phi$, skyrmion number $q$ and helicity $\gamma_{helicity}$ as

$$\Theta(\mathbf{r}) = \Theta(r) = 2\arctan\left(\frac{\sinh(r/w)}{\sinh(R/w)}\right), \quad (5)$$

$$\Phi(\mathbf{r}) = \Phi(\varphi) = q\varphi + \gamma_{helicity}$$

with boundary conditions of $\Theta(0)=0$ and $\Theta(\infty)=\pi$, where $(r,\varphi)$ are the polar coordinate system: $\mathbf{r}=(x, y)=(r\cos\varphi, r\sin\varphi)$ [2,9]. In this paper, the skyrmion number $q=1$ and $\gamma_{helicity}=0$ or $\pi$ for the Néel skyrmions.

# References

1. Skyrme, T. H. R. A unified field theory of mesons and baryons. *Nuclear Phys*. **31**, 556–569 (1962).

2. Nagaosa, N. & Tokura, Y. Topological properties and dynamics of magnetic skyrmions. *Nat. Nanotechnol.* **8**, 899-911 (2013).

3. Zhang, S., Baker, A. A., Komineas, S. & Hesjedal, T. Topological computation based on direct magnetic logic communication. *Sci. Rep*. **5**, 15773 (2015).

4. Zazvorka, J. *et al*. Thermal skyrmion diffusion used in a reshuffler device. *Nat. Nanotechnol.* **14**, 658-661 (2019).

5. Jibiki, Y. *et al*. Skyrmion Brownian circuit implemented in a continuous ferromagnetic thin film. arXiv. 1909.10130 (2019).

6. Nozaki, T. *et al.* Brownian motion of skyrmion bubbles and its control by voltage applications. *Appl. Phys. Lett.* **114**, 012402 (2019).

7. Thiele, A. A. Steady-state motion of magnetic domains. *Phys. Rev. Lett*. **30**, 230 (1973).

8. Belavin, A. A. & Polyakov, A. M. Metastable states of two-dimensional isotropic ferromagnets. *JETP Lett.* **22**, 245 (1975).

9. Wang, X. S., Yuan, H. Y. & Wang, X. R. A theory on skyrmion size, *Comms. Phys*. **1**, 31 (2018).

10. A. Vansteenkiste *et al*, The design and verification of MuMax3. *AIP Advances* **4**, 107133 (2014).

11. Dierckx, P. An algorithm for smoothing, differentiation and integration of experimental data using spline functions. *J. Comp. Appl. Maths* **1,** 165-184 (1975).

12. Xia, H. *et al.* Skyrmion motion driven by the gradient of voltage-controlled

magnetic anisotropy. *J. Magn. Magn. Mater*. **458**, 57–61 (2018).

13. Ang, C. C. I., Gan, W. & Lew, W. S. Bilayer skyrmion dynamics on a magnetic anisotropy gradient. *New J. Phys.* **21**, 043006 (2019).

14. Shen, L. *et al.* Dynamics of the antiferromagnetic skyrmion induced by a magnetic anisotropy gradient. *Phys. Rev. B*. **98**, 134448 (2018).

15. Toscano, D. *et al.* Suppression of the skyrmion Hall effect in planar nanomagnets by the magnetic properties engineering: Skyrmion transport on nanotracks with magnetic strips. *J. Magn. Magn. Mater.* **504,** 166655 (2020).

16. Büttner, F. *et al.* Dynamics and inertia of skyrmionics spin structures. *Nat. phys.* **11**, 225-228 (2015).

17. Martinez, J. C. & Jalil, M. B. A. Mass of a skyrmion under a driving current. *J. Magn. Magn. Mater.* **424**, 291–297 (2017).

18. Saitoh E., Miyajima, H., Yamaoka, T. & Tatara, G. Current-induced resonance and mass determination of a single magnetic domain wall. *Nature.* **432**, 203–206(2004).

19. Yan, M., Andreas, C., Kákay, A., Sánchez, F. G. & Hertel, R. Fast domain wall dynamics in magnetic nanotubes: Suppression of Walker breakdown and Cherenkov-like spin wave emission. *Appl. Phys. Lett*. **99**, 122505 (2011).

Figures and Captions

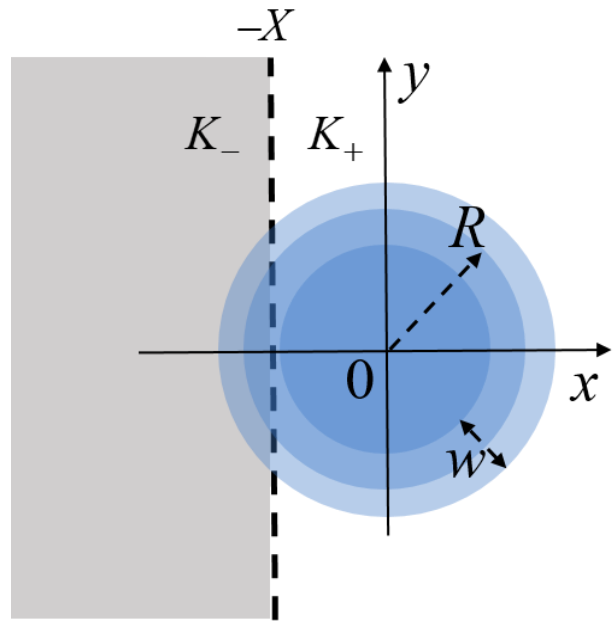

**Fig. 1. Coordinates of a skyrmion and a wall.** The magnetic anisotropy of the shadowed region is $K_-$ whose value is higher than $K_+$.

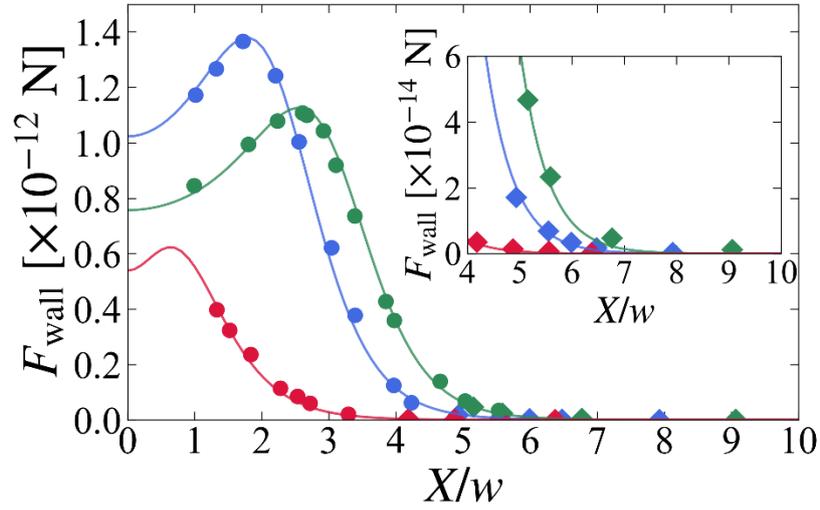

**Fig. 2. The force between skyrmions and walls depending on the relative positions X.** The solid curves and the dots and diamonds represent the theory and the simulations. The colors, blue, green, and red, indicate different skyrmions whose radii and the domain wall widths ($R$, $w$) are (25.55 nm, 11.35 nm), (24.60 nm, 8.05 nm) and (7.4 nm, 7.9 nm). The dots are obtained from the balance between $F_{wall}$ and the Magnus force **G**×**v**, and diamonds from between $F_{wall}$ and the gradient force $F_g$. The inset shows the zoom-in view.